\documentclass{egpubl}
\usepackage{eg2023}
\usepackage{amssymb}
\usepackage{multirow}
\usepackage{lineno}

\ConferencePaper        
\usepackage[T1]{fontenc}
\usepackage{dfadobe}  
\bibtexVersion
\BibtexOrBiblatex
\electronicVersion
\PrintedOrElectronic

\ifpdf \usepackage[pdftex]{graphicx} \pdfcompresslevel=9
\else \usepackage[dvips]{graphicx} \fi

\usepackage{egweblnk} 

\usepackage{defines}

\title[In-the-wild Material Appearance Editing using Perceptual Attributes]%
      {In-the-wild Material Appearance Editing using Perceptual Attributes}

\author[Subias \& Lagunas]
{\parbox{\textwidth}{\centering J. Daniel Subias$^{1}$\orcid{0000-0002-5480-7462}
        and M. Lagunas$^{2}$
        \orcid{0000-0003-0838-1795}}
        \\
{\parbox{\textwidth}{\centering $^1$Universidad de Zaragoza, I3A, Spain  \\$^2$Amazon, Spain}
}
}
\begin{document}
\teaser{
 \includegraphics[width=\linewidth]{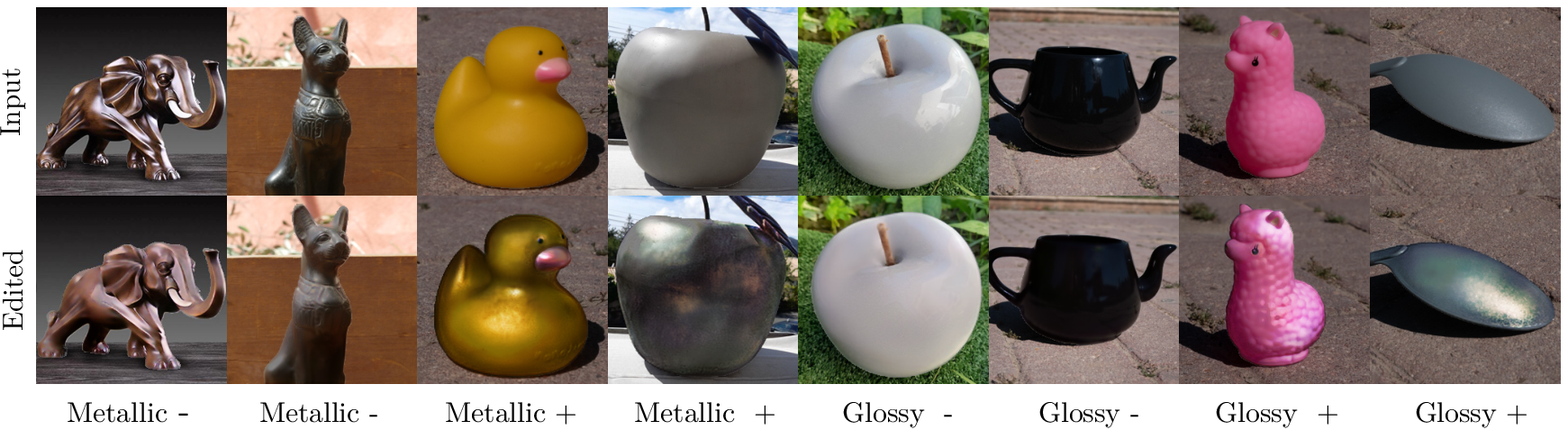}
 \centering
  \caption{\label{teaser} 
Our approach enables intuitive appearance editing of high-level perceptual attributes. Our framework takes as an input a single image of an object (top) and produces high-quality edits of material attributes such as \textit{glossy} or \textit{metallic}, while preserving the geometrical structure and details (bottom). The``+'' and ``-'' indicate whether the target high-level perceptual attribute is increased or decreased.}
\label{fig:teaser}
}
\maketitle
\begin{abstract}
   Intuitively editing the appearance of materials from a single image is a challenging task given the complexity of the interactions between light and matter, and the ambivalence of human perception. This problem has been traditionally addressed by estimating additional factors of the scene like geometry or illumination, thus solving an inverse rendering problem and subduing the final quality of the results to the quality of these estimations. We present a single-image appearance editing framework that allows us to intuitively modify the material appearance of an object by increasing or decreasing high-level perceptual attributes describing such appearance (e.g., \textit{glossy} or \textit{metallic}). Our framework takes as input an in-the-wild image of a single object, where geometry, material, and illumination are not controlled, and inverse rendering is not required. We rely on generative models and devise a novel architecture with Selective Transfer Unit (STU) cells that allow to preserve the high-frequency details from the input image in the edited one. To train our framework we leverage a dataset with pairs of synthetic images rendered with physically-based algorithms, and the corresponding crowd-sourced ratings of high-level perceptual attributes. We show that our material editing framework outperforms the state of the art, and showcase its applicability on synthetic images, in-the-wild real-world photographs, and video sequences.
\end{abstract}  
\section{Introduction}

Humans are visual creatures, visual data play a key role in the way we perceive and understand the world around us. We are able of recognizing materials, understanding their appearance, and reasoning about other physical properties effortlessly, just briefly looking at them. The visual appearance of a material --- whether it appears \textit{glossy}, \textit{metallic}, or \textit{rough} --- is one of the key properties that determine how we manipulate and interact with objects. However, such visual appearance is formed by a complex multidimensional interaction involving the material properties themselves, but also other confounding factors like geometry or illumination, that can affect our perception. Unfortunately, such underlying perceptual process is not yet completely understood~\cite{anderson11CB,fleming14VR,maloney10JoV}.

Given the large amount of dimensions that our perception involves, many works have focused on editing material appearance by estimating the known parameters of the scene (material, geometry, and illumination), thus solving an inverse rendering problem where the material is later modified~\cite{bati21ACMTG,yang16SIG}.
However, this approach faces several problems: noise in the estimation of scene parameters can yield uncanny editing results. Moreover, the user should understand the intricacies of the material formulation to be able to edit. Techniques and hardware to acquire material appearance are gaining accuracy, speed, and efficiency~\cite{nielsen15ACMTG,aittala15ACMTG,dupuy18ACMTG}; creating a data-driven shift for appearance editing techniques~\cite{serrano16ACMTG,zheng22ACMTG,zsolnai20CGF}
. These methods still present similar problems as using inverse rendering. There exists a disconnection between the mathematical formulation of material appearance and human-friendly parameters. The captured data is machine friendly but not human friendly. Editing material appearance given a single in-the-wild photograph, and using intuitive perceptual attributes is therefore a challenging task that remains to be solved. 

In this work, we present an image-based framework that does not rely on any physically-based rendering but instead modifies directly the material appearance in the image space. It takes a certain image of an object as input and modifies the appearance based on varying the value of the desired high-level perceptual attribute describing appearance (see Figure \ref{fig:teaser}). Recently, the popularization of generative deep learning models had allow us to design data-driven frameworks for image editing~\cite{liu19CVPR,he19IEETIP,lample17ANIPS}. Since the image cues that drive the perception of such attributes can not be captured in a few image statistics, we rely on such generative neural networks to learn their relationship with material appearance and generate novel edited images~\cite{storrs21nature, lagunas21JoV}. 

We devise a generative architecture that allows to intuitively edit material appearance just from a single, in-the-wild, image and given the value of the target perceptual attribute that we want to manipulate. 
To train our framework, a first approach might be collecting pairs of images the originals and the edited ones, where the edited examples were produced given a target high-level perceptual attribute value. This approach is not only tedious but also hinders the training process. We follow the work from Delanoy et al.~\cite{delanoy22CGF}, which is based on training a two-step generative framework using a wide dataset composed of a large variety of images, paired with high-level perceptual ratings obtained through user studies. However, this method needs additional geometry information (as a normal map) of the target object. We thus devise an encoder-decoder architecture that allows us to send the relevant high-frequency information of the object's shape from the encoder to the decoder thanks to the STU cell~\cite{liu19CVPR} and a novel loss function. Thus, removing the need for the normal map as the input and not subduing our results in in-the-wild photographs to having good estimations of the normal map.

We trained two version of our framework focusing on two high-level perceptual attributes that are both common and easy to understand: \textit{metallic} and \textit{glossy}. We evaluate the consistency of our edits on a wide variety of scenes with different illuminations, materials, and geometries; and compare our results with the work from Delanoy et al.~\cite{delanoy22CGF}. We observe that our method, despite resorting to simpler input and having a simpler architecture, obtains more realistic material appearance edits of the input image. We also assess the temporal consistency by editing video sequences composed of frames rendered using unseen illumination, geometry, and materials. We observe that our framework is capable of producing coherent outputs even when the additional temporal dimension is included. 
\section{Related Work}

\subsection{Visual Perception}
Understanding how our visual system interprets our world is a longstanding goal in fields like computer graphics or applied optics~\cite{nieves21JoI}, that is yet to be understood~\cite{fleming01,fleming03JoV}. Our visual perception of an object is guided by its material properties; but, also involves factors such as geometry~\cite{vangorp07SIG,havran16CGF}, light conditions~\cite{ho06JoV,krivanek10ACMTG,chadwick15VR} or motion~\cite{doerschner11CB,mao19ACMSAP}. To reduce the dimensionality of the problem, previous work have focused on understanding single, high-level appearance properties like \textit{glossiness}~\cite{pellacini00SIG,wills09ACMTG,chadwick15VR}, translucency~\cite{gkioulekas13ACMTG,xiao20JoV,gkioulekas15CVPR} and softness~\cite{schmidt20JoV,cavdan21bioR}; or draw inspiration from artists' implicit understanding of the key visual cues that guide visual perception~\cite{di19JoV,delanoy21JoV}. Recent works suggest that material perception may be driven by complex non-linear statistics better approximated by highly non-linear models such as neural networks~\cite{fleming19COBS,storrs21nature,delanoy20SIG,lagunas19SIG}. Inspired by this, we propose a deep-learning-based framework for material appearance editing that relies on images paired with human judgments about high-level perceptual attributes to be trained.

\subsection{BRDF-based Material Editing}

Editing material appearance is a complex task since there is a disconnection between our perception and materials' physical properties~\cite{fleming13JoV,thompson11CRC,chadwick15VR}. Non-parametric models such as SVBRDFs are hard to edit. Different approaches have been proposed, such as inverse shading trees~\cite{lawrence06ACMTG}, procedural models~\cite{hu22ACMTG}, or using deep-learning techniques~\cite{deschaintre20CGF}. Several perceptually-based frameworks have been proposed to provide intuitive controls over parametric~\cite{james01SPIE,pellacini00SIG} and non-parametric appearance models~\cite{serrano16ACMTG,mylo17PCV,hu20CGF}. Unfortunately, these models only capture material properties, while leaving out other scene parameters that also drive our perception of material appearances, such as geometry and illumination~\cite{lagunas21JoV}. 
 Another line of work proposes an intrinsic decomposition of the scene to manipulate materials~\cite{barron15IEETPA,yu19CVPR}. More recently, NeRF~\cite{mildenhall20ECCV} based approaches for such decomposition are allowing for unprecedented levels of realism~\cite{boss21ICCV,zhang21ACMTG,srinivasan21CVPR}. However, these methods only provide a new material definition that can later be used in a specific 3D scene but do not allow to edit it. Our work presents an intuitive framework that directly uses in-the-wild images and is capable of editing the appearance of materials using high-level attributes such as \textit{glossiness}. 

\subsection{Image-based Material Editing}

Image-based material editing attempts to manipulate the pixel values directly into the image space. Several interactive frameworks have been proposed to provide users more intuitive controls over the target editing regions of the image, selecting them just from a few strokes~\cite{pellacini07SIG,an08SIG,dong11SIG}. The work from Khan et al.~\cite{khan06ACMTG} proposes a single-image editing framework exploiting our tolerance to certain physical inaccuracies. Such approach was extended later to relight paintings~\cite{lopezmoreno10ACMNPAR}, consider global illumination and caustics~\cite{gutierrez08ACMTG}, or weathering effects~\cite{xuey08CGF}.
More recent work, supported by the success of deep-learning-based methods, introduces editing approaches by factoring the image into shape, material, and reflectance~\cite{liu17ICCV,rematas16CVPR,maximov19ICCV}.
Splitting an image into separate components like frequency bands~\cite{boyadzhiev15ACMTG} or shading and reflectance~\cite{garces12CGF} has been a standard practice for image manipulation.
Instead, we propose a generative-based editing framework without the need to decompose the input image, learning to edit directly the visual cues that drive our perception of high-level attributes. 
Generative Adversarial Networks (GANs)~\cite{goodfellow14ANIPS} have been proposed to edit face attributes (i.e., hair color or gender) through the latent space~\cite{lample17ANIPS,he19IEETIP,liu19CVPR}.
Our framework works instead in the complex problem of material appearance, where confounding factors like geometry or illumination have a direct impact in our perception.
Closer to ours, Delanoy et al.~\shortcite{delanoy22CGF} introduce a generative framework for intuitive appearance editing using high-level perceptual attributes. However, their method requires two inputs: the image, and its normal map which yields non-photorealistic results for in in-the-wild images where the normal map is estimated. We devise a novel generative architecture that allows to keep the high-frequency information from the geometry of the input images, thus removing the need for the normal map as the input while obtaining superior performance.

\section{Our Framework}

This section describes our proposed framework for single-image appearance editing. We introduce our goal (Section~\ref{sec:goal}), describe the mathematical model of the Selective Transfer Units (STU) cells  (Section~\ref{sec:stu}), and, explain the different modules that make our model architecture is built upon (Section~\ref{sec:architecture}).

\subsection{Goal and Overview} 
\label{sec:goal}
Our goal is to generate an image $\mathbf{y}$ whose material appearance we want to edit from an input image $\mathbf{x}$ and a value $\mathbf{a t t}_t \in [0,1]$ of the target high-level perceptual attribute to edit (e.g., \textit{glossy} or \textit{metallic}). The target edited image $\mathbf{y}$ depicts the same object as $\mathbf{x}$ and elicits a visual appearance according to the value of the high-level perceptual attribute $\mathbf{a t t}_t$. For instance, more \textit{glossy} if $\mathbf{a t t}_t$ is closer to $1$ and less if closer to $0$. 
To achieve it, we introduce a novel framework that relies on an encoder-decoder architecture $G$ that encodes the image $\mathbf{x}$, and manipulates the latent space $\mathbf z$ together with the target attribute $\mathbf{a t t}_t$ to generate the edited image $\mathbf{y}$. A high-level overview of our framework is shown in Figure~\ref{fig:architecture_stu}.

\subsection{Selective Transfer Units}
\label{sec:stu}

When using encoder-decoder architectures, it is common practice to send information between the encoder and decoder by using skip-connections~\cite{ronneberger15Spr}. This allows us to keep high-frequency details in the generated image that are lost otherwise~\cite{kingma13arxiv,higgins17ICLR}. However, in image editing tasks that manipulate the latent space, adding skip connections hampers the editability of the model, where only the input image can be reconstructed~\cite{choi18CVPR,he19IEETIP,delanoy22CGF}. To bridge this gap, we send information from the encoder to the decoder by selectively removing unnecessary data using Selective Transfer Units (STU) memory cells~\cite{liu19CVPR}.

The STU architecture, illustrated in Figure~\ref{fig:architecture_stu}, is a variant of the GRU~\cite{cho14ACLEM,chung14CoRR} and allows encoder-decoder architectures to keep the relevant information of the input image in the edited output when manipulating the latent space $\mathbf z$. Given the feature map of the $l^{t h}$ encoder layer denoted by $\mathbf{f}^l_{enc}$, the STU cell outputs an edited feature map $\mathbf{f}^l_{t}$, as is shown in Figure~\ref{fig:architecture_stu}. 
Each STU cell receives information from the previous cell via a feature map $\hat{\mathbf{s}}^{l+1}$ (also called hidden state), which also contains information of the target attribute $\mathbf{att}_t$. The STU updates its internal hidden state denoted as $\mathbf{s}^{l}$ and sends this to the next STU cell. For further information about the mathematical formulation of STU cells refer to Appendix~\ref{apx:stu}.

\subsection{Network Architecture}
\label{sec:architecture}

Our framework is a GAN-like model composed of a generator $G$ and a discriminator $D$ that is only used during training. 
Our goal is to leverage $G$ to edit an in-the-wild input image $\mathbf x$ according to a target high-level perceptual attribute describing appearance $\mathbf{a t t}_t$ to generate an edited image $\mathbf y$ where
\begin{equation}
    \mathbf{y} = G(\mathbf{x}, \mathbf{a t t}_t).
\end{equation}
The generator $G$ is composed of an encoder module $G_{e n c}$ that encodes the input image to a latent vector $\mathbf z$ and 
a decoder module $G_{d e c}$ that generates the edited image,
both with the same number $n$ of layers. 

\begin{figure*}[t!]
    \centering
    \includegraphics[width=\linewidth]{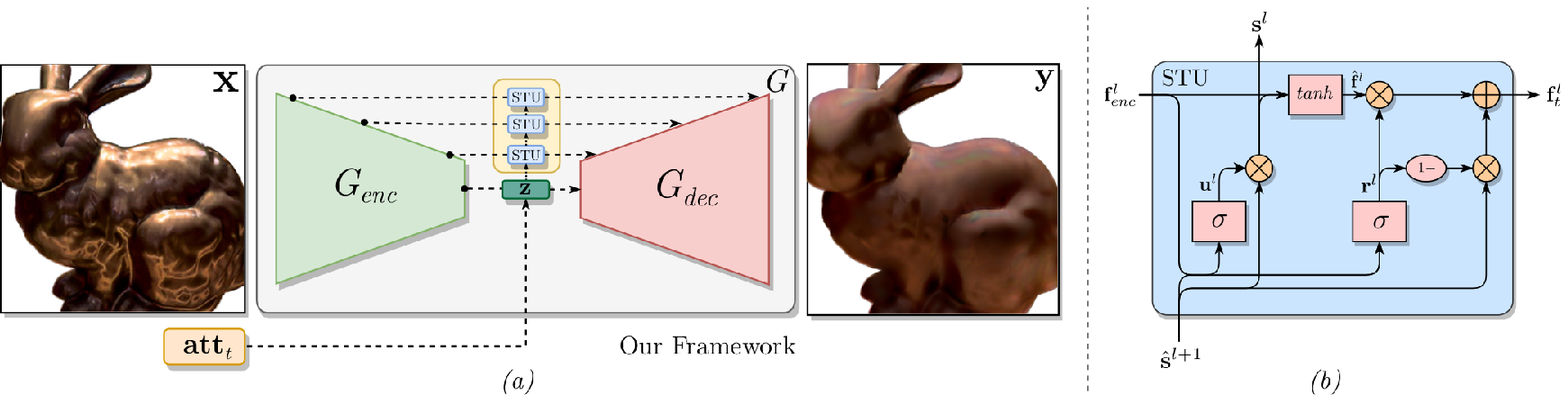}
    \caption{ \label{fig:architecture_stu}
    (a) High-level overview of our framework. Our generator $G$ is composed of an encoder $G_{e n c}$ and a decoder $G_{d e c}$. It is capable of editing the input image $\mathbf x$ according to the target attribute $\mathbf{att}_t $ to generate the edited image $\mathbf y$. (b) The architecture of a single STU cell. As an input, it takes the feature map of the current layer $\mathbf{f}^l_{enc}$ and the hidden state of the previous cell $\hat{\mathbf{s}}^{l+1}$. It outputs the updated hidden state $\mathbf{s}^l$ and feature map $\mathbf{f}_t^l$.}
\end{figure*}

The encoder $G_{e n c}$ compresses the image $\mathbf{x}$ in a latent code $\mathbf{z}$, while storing the set of features maps $\mathbf{f}= \{\mathbf{f}^1_{enc},\mathbf{f}^2_{enc},...,\mathbf{f}^{n-1}_{enc}\}$, generated by its convolutional layers. The latent code $\mathbf{z}$ corresponds to the feature map generated by the last convolutional layer such that $\mathbf{z} = \mathbf{f}^n_{enc}$. The decoder $G_{d e c}$ reconstructs the edited image $\mathbf{y}$ from the latent code $\mathbf{z}$ concatenated with the target high-level perceptual attribute $\mathbf{att}_t$. 
To keep the high frequencies from the input in the edited image $\mathbf{y}$, we send the feature maps $\mathbf{f}$ via skip-connections. However, performing this process without additionally processing the information in the features affects the editing ability of the generator $G$. Thus, we introduce an STU cell in each skip-connection where $G_{st}$ denotes the set of STU cells of the framework, and $G_{st}^l$ corresponds to the STU cell of the $l^{t h}$ layer.
The STU cells edit the set of feature maps $\mathbf{f}$ from the input, generating a set of edited ones $\mathbf{f}_t  = \{ \mathbf{f}_t^{1}, \mathbf{f}_t^{2} , ..., \mathbf{f}_t^{n-1}\}$. Since the STU cell of the deepest skip connection $G_{st}^{n-1}$ does not receive a hidden state $\hat{\mathbf{s}}$ from another STU cell, $G_{st}^{n-1}$ takes as input the latent code $\mathbf{z}$ concatenated with the target attribute $\mathbf{a t t}_t$ to edit the feature map $\mathbf{f}^{n-1}_{enc}$, as is shown in Figure~\ref{fig:architecture_stu}. For further details of our architecture refer to Appendix~\ref{apx:architecture_details}.

\section{Learning to Edit Material Appearance}

Training GAN-like models is a complex task that requires careful tuning of the hyperparameters. 
We first describe the dataset of images, with paired crowd-sourced ratings of high-level perceptual attributes, that we used to train our framework (Section~\ref{sec:training_dataset}). Then we introduce the loss function that allows us to faithfully edit material appearance according to the target perceptual attribute (e.g., \textit{glossy}) (Section~\ref{sec:losses}), and last, we describe the technical details of the training process (Section~\ref{sec:training_details}).

\subsection{Training Dataset}

\label{sec:training_dataset}

\begin{figure}[t!]
  \centering
  \includegraphics[width=\linewidth]{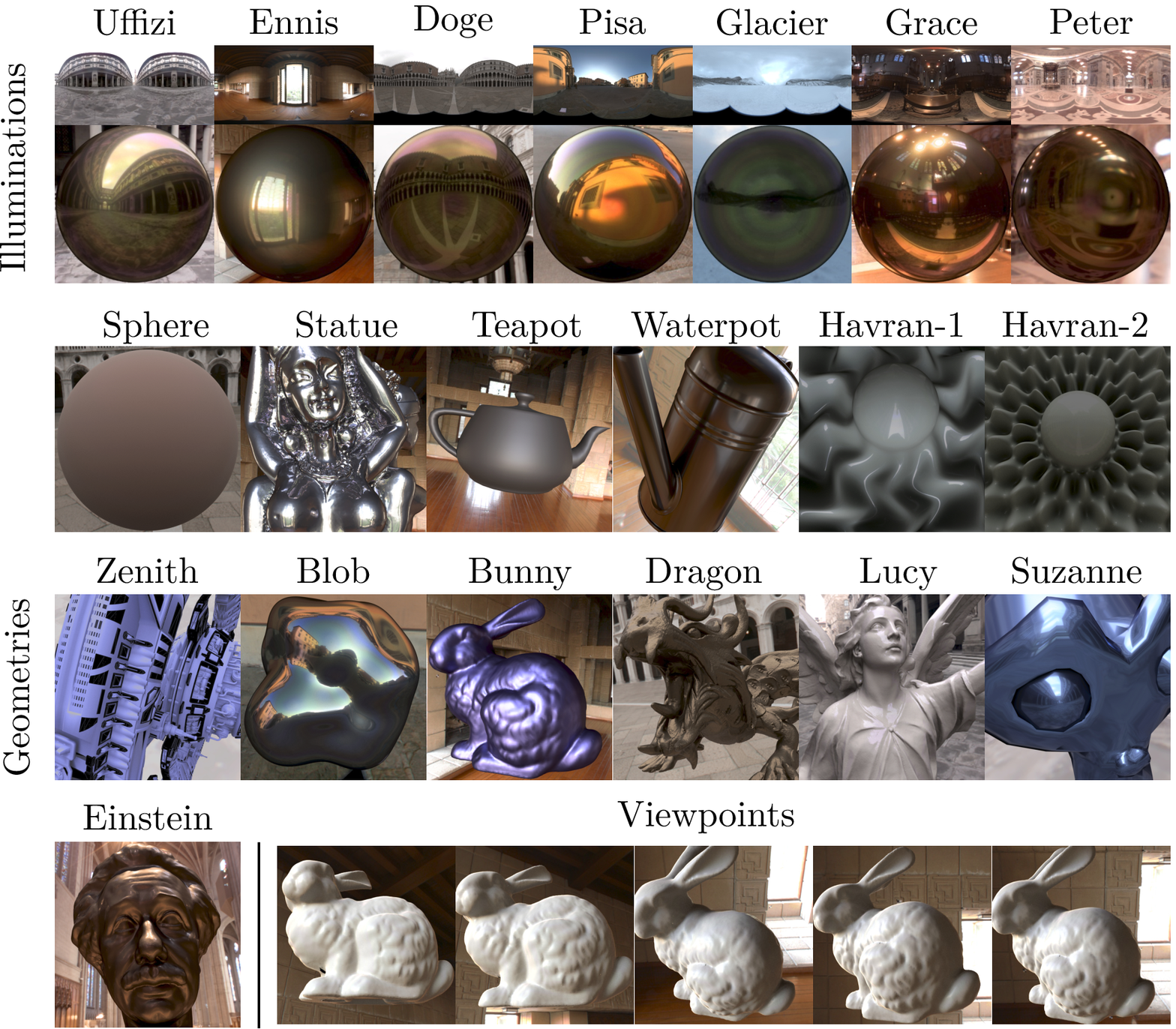}
  \caption{\label{fig:training_data}
   Seven illuminations present in the training dataset and corresponding rendered spheres with the \textit{brass} material (top). Thirteen scenes of the geometries present in the training dataset,  and five viewpoints of the bunny geometry rendered using the nylon material and \textit{Ennis} illumination (bottom).}

\end{figure}

Generating ground-truth data from analytical BRDF models is not a robust approach to training our framework, since their parameters are not aligned with our perception~\cite{ngan06SR,wills09ACMTG}. In addition, our framework would learn to edit based on variations in a physical parameter (e.g., roughness) and not on variations in our perception. Thus, we leverage the dataset of Delanoy et. al~\cite{delanoy22CGF}, designed for material appearance perception tasks.
This dataset based on the Lagunas et al. dataset~\cite{lagunas19SIG} contains renderings of 13 different geometries, illuminated by 7 captured real-world illuminations~\cite{debevec}. Renders have been made with the physically-based renderer Mitsuba~\cite{Mitsuba} using 100 different BRDFs from the Merl dataset~\cite{matusik03ACMTG}. For each combination of material $\times$ shape $\times$ illumination, 5 different images with slight variations in the viewpoint (randomly sampled within a 45 degrees cone around the original viewpoint) have been rendered, as is shown in Figure~\ref{fig:training_data}. The dataset has 45,500 images (13 geometries $\times$ 100 materials $\times$  7 illuminations $\times$ 5 views). 

Each scene is described by a set of high-level perceptual attributes (i.e., \textit{plastic}, \textit{rubber}, \textit{metallic}, \textit{glossy}, \textit{bright}, \textit{rough}, and the strength and sharpness of \textit{reflections}) rated on a normalized Likert scale, from 0 to 1. A total of 2,600 paid subjects participated in the study, each of them rating 15 different random images. The final dataset gathers a perceptual rating per viewpoint. To reduce the presence of noise and outliers, we use the median value pooled over the 5 viewpoints and the shape. Then the attribute values are different for each material and illumination.

\paragraph*{Data Augmentation} To have more diversity in the images and help our model generalize better, we add a data augmentation pipeline. First, we resize the images to $512 \times 512$ px; then we flip and rotate them 90 degrees randomly, and we finally crop them to $480 \times 480$ px. In addition, to reduce the bias on the colors BRDF we represent images on the HSV color space and make a shift in the Hue and Saturation channels. Finally, the input is resized to a size of $256 \times 256$ px. to remove the background. Within each input image, the background is removed by applying a mask of the object's silhouette. Each image in the training dataset is paired with a high-level perceptual attribute $\mathbf{att}_s \in [0,1]$; however, with our framework, we want to edit the input images with a different target attribute $\mathbf{att}_t$. During training, we sample $\mathbf{att}_t$ randomly using a uniform distribution $\mathbf{U}([0,1])$ allowing the generator $G$ to produce edited images different from the input to trick the discriminator $D$.

\subsection{Loss Functions}
\label{sec:losses}

\begin{figure}[t!]
  \centering
  \mbox{} \hfill
  \centering
    \includegraphics[width=\linewidth]{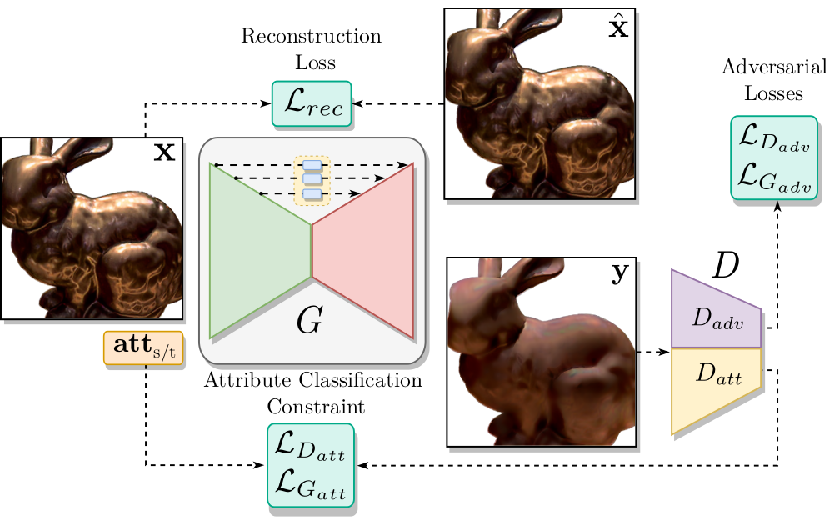}
    \caption{\label{fig:g_schema} 
    Training scheme of our framework. The gray block represents our generator $G$, and the discriminator branches $D_{a d v}$ and $D_{a t t}$ are illustrated by the purple and yellow blocks, respectively. Pointed arrows denote the parameters used as input for the training losses (Section~\ref{sec:losses}).}
\end{figure}

We adopt the adversarial training proposed by He et al.~\cite{he19IEETIP} and introduce a GAN model where the discriminator $D$ has two branches $D_{a d v}$ and $D_{a t t}$. $D_{a d v}$ consists of five convolution layers to predict whether an image is fake (edited) or real. $D_{a t t}$ shares the convolution layers with $D_{a d v}$ and, instead, predicts the high-level attribute value $\mathbf{att}_t $.  Figure~\ref{fig:g_schema} shows a high-level scheme of our framework during training.

\paragraph*{Adversarial Losses} 
Since we do not know what are the ground-truth edited results,
we use an adversarial loss~\cite{goodfellow14ANIPS} aiming to generate edited results indistinguishable from real images. GANs are complex to train, partially due to the instability of the loss function proposed in the original formulation~\cite{wasserstein17PMLR}. Thus, we rely on the WGAN-GP~\cite{gulrajani17ANIPS} loss to alleviate such a problem. The discriminator $D$ is trained to give a high score to real images and a low score to generated ones, aiming to disambiguate them, and tries to minimize the adversarial loss defined as:

\begin{equation}
    \mathcal{L}_{D_{adv}} = \mathbb{E}_\mathbf{x} \left[ D_{adv} (\mathbf{x})\right] - \mathbb{E}_\mathbf{y} \left[D_{adv}(\mathbf{y}) \right]+ \mathcal{L}_{G P}.
    \label{eq:disc_loss}
\end{equation}

Formally, $D$ should be a 1-Lipschitz continuous function. To keep this constraint, WGANs~\cite{wasserstein17PMLR} introduces a gradient penalty term defined as follows: 

\begin{equation}
    \mathcal{L}_{G P} = \lambda_1 \mathbb{E}_\mathbf{\hat{x}} \left[ (|| \nabla_{\mathbf{\hat{x}}} D_{adv} (\mathbf{\hat{x}})||_2 - 1)^2 \right].
    \label{eq:gradient_penalty}
\end{equation}

Where $\lambda_1$ is the gradient penalty weight. The generator $G$ is trained such that the discriminator $D$ believes that generated images are real, giving them a high score. Therefore the generator's adversarial loss is given by:

\begin{eqnarray}
    \mathcal{L}_{G_{adv}} = \mathbb{E}_{\mathbf{y}}\left[ D_{adv} (\mathbf{y})\right].
\end{eqnarray}

\paragraph*{Attribute Manipulation Losses} Even though the ground truth is missing, the editing result has to elicit a visual impression according to the target perceptual attribute $\mathbf{a t t}_t$. Therefore, we introduce an attribute classifier $D_{a t t}$ which learns to predict the attribute values $\mathbf{a t t}_s$ from the images that belong to the training dataset. Since we have to compare the distance from the predicted attributes by $D_{a t t}$ to $\mathbf{a t t}_s$, the following attribute manipulation loss is computed and optimized by $D_{a t t}$ during training:

\begin{equation}
    \mathcal{L}_{D_{a t t}} = || \mathbf{a t t}_s - D_{a t t} (\mathbf{x})||_1.
\end{equation}

The generator $G$ should produce images with similar looks to real ones to trick $D$, so its edits must be consistent with the target perceptual attributes $\mathbf{a t t}_t$, minimizing the following distance:

\begin{equation}
    \mathcal{L}_{G_{a t t}} = || \mathbf{a t t}_t - D_{a t t} (\mathbf{y})||_1.
\end{equation}

\paragraph*{Reconstruction Loss} Each image in the training set has crowd-sourced ratings of high-level perceptual attributes $\mathbf{a t t}_s$, describing their appearance. Thus, if we ask our generator $G$ to edit $\mathbf{x}$ according to the attribute $\mathbf{a t t}_s$, it should generate an edited image $\hat{\mathbf{x}}$ similar to the input. We minimize the following L1-norm during training:

\begin{equation}
    \mathcal{L}_{rec} = || \mathbf{x} - \hat{\mathbf{x}}||_1.
\end{equation}

\paragraph*{Final Loss} Taking the above losses into account, the objectives to train the discriminator $D$ and generator $G$ can be formulated as:

\begin{eqnarray}
    \min_D \mathcal{L} & = & -\mathcal{L}_{D_{a d v}} + \lambda_2 \mathcal{L}_{D_{a t t}}, \label{eq:final_loss_d}\\
    \min_G \mathcal{L} & = & -\mathcal{L}_{G_{a d v}} + \lambda_3 \mathcal{L}_{G_{a t t}} + \lambda_4 \mathcal{L}_{r e c}.
    \label{eq:final_loss_g}
\end{eqnarray}

Where $\lambda_2$, $\lambda_3$, and $\lambda_4$ are the model trade-off parameters, that are tunned to yield values of the individual losses in a similar order of magnitude. Both $G$ and $D$ will try to minimize their objective function, however, once $G$ has learned to trick $D$, the loss function of the latter will tend to increase until both models reach a stable equilibrium.

\subsection{Training Details} 
\label{sec:training_details}

We train the model using the ADAM optimizer~\cite{kingma14arxiv} with $\beta_1 = 0.5 $ and $\beta_2 = 0.999$. The learning rate $\eta$ is $2 \times 10^{-4}$ for both $G$ and $D$. The trade-off parameters are $\lambda_2 = 50$, $\lambda_3 = 100$ and $\lambda_4 = 1000$ for the Equations~\ref{eq:final_loss_d} and~\ref{eq:final_loss_g}, and $\lambda_1 = 10$ for Equation~\ref{eq:gradient_penalty} (see Appendix~\ref{apx:architecture_details} for further details). At each training step, the discriminator iterates 7 times while the generator updates its parameters once. The memory module $G_{st}$ is composed of 4 skip connections with STU cells between the encoder and decoder. All the experiments are computed using the PyTorch~\cite{paszke19ANIPS} library with cuDNN 7.1, running on an Nvidia GeForce RTX 3090 GPU. In total, training the whole framework takes two days. 

\section{Evaluation and Results}

We evaluate our framework on a set of synthetic images and real photographs that have not been seen by the model during training. We start by describing the evaluation dataset(Section~\ref{sec:test_dataset}); validate the design choices of our framework with a series of ablation studies (Section ~\ref{sec:ablation_studies}); demonstrate the robustness of the proposed method by comparing the obtained results with varying geometry, illumination, or material (Section~\ref{sec:consistency}); and finally compare our method to the state of the art, obtaining better performance (Section~\ref{sec:comparison}). 

\subsection{Evaluation Dataset}
\label{sec:test_dataset}

\begin{figure}[t!]
  \centering
    \mbox{} \hfill
    \includegraphics[width=\linewidth]{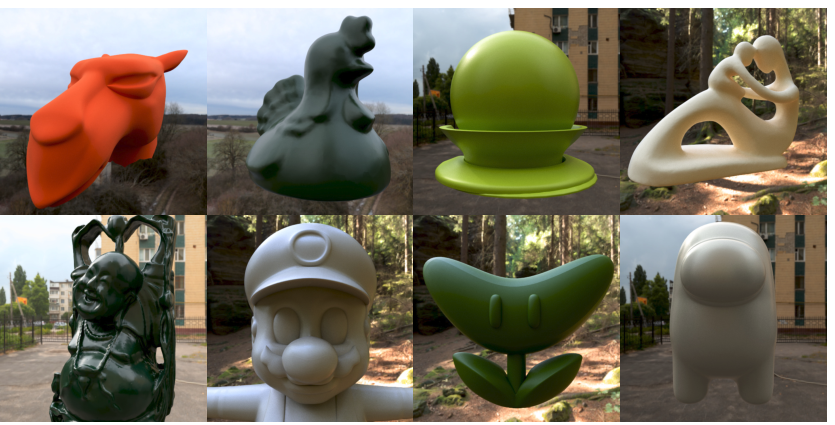}
    \caption{\label{fig:synthetic_samples} Synthetic dataset used to evaluate our framework. The images show eight geometries rendered with illumination and materials never seen in the training process.}
\end{figure}

\begin{figure}[t!]
  \centering
    \mbox{} \hfill
    \includegraphics[width=\linewidth]{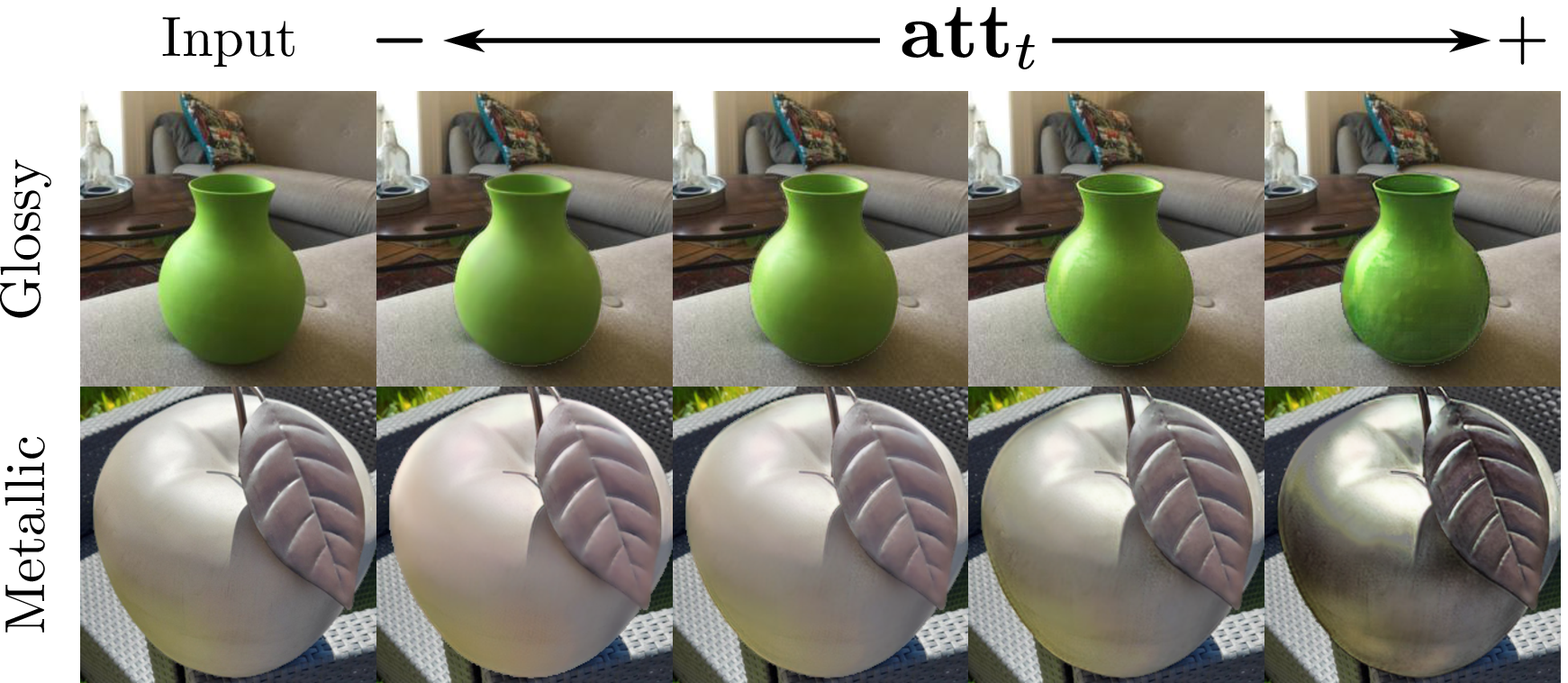}
    \caption{\label{fig:editing_metallic} Editing results by varying the \textit{metallic} or \textit{glossy} attributes. The ``+'' and ``-'' indicate whether the target high-level perceptual attribute is increased or decreased.}
\end{figure}

To evaluate our framework we leverage the synthetic dataset used by Delanoy et al.~\cite{delanoy22CGF}, containing scenes with shapes and materials never seen during training. They have been rendered using eight shapes collected from online sources, four illuminations obtained from HDRI-Haven~\shortcite{hdrihaven}, and eight materials coming from Dupuy and Jakob’s database~\cite{dupuy18ACMTG}. Figure~\ref{fig:synthetic_samples} shows a representative subset of the synthetic scenes. We also test our framework's performance with real-world photographs downloaded from online catalogs of decorative items and with in-the-wild mobile photos taken by us. We masked the object of interest using the online API Kalideo~\cite{kalideo}. Figure~\ref{fig:editing_metallic} shows material appearance editing results using our framework sampling different values for the target attribute $\mathbf{a t t_t}$. We can see the consistency of our edits when the attribute varies for in-the-wild photographs.

\subsection{Ablation Studies}
\label{sec:ablation_studies}
\begin{figure}[t]
    \centering
    \mbox{} \hfill
    \includegraphics[width=\linewidth]{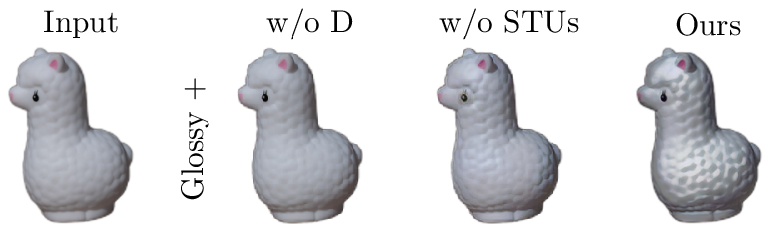}
    \caption{\label{fig:ablation_modules} Ablation studies where we analyze the impact of the different components of our framework. From left to right: input image, edited image without the discriminator $D$ during training, edited image without using the STU cell, and our framework. We can see how by not using the discriminator $D$ the framework is not able to edit and only reconstructs the input image, and using only skip connections, without the STU, hampers the ability to edit. The ``+'' indicates an increase of the target attribute.
    }
\end{figure}

We evaluate our design choices via a series of ablation studies. Our framework is based on a generative architecture that uses a discriminator $D$ during training and STU cells~\cite{liu19CVPR} to obtain more realistic edited images while keeping the relevant high-frequency details of the input image. 
The model without the discriminator $D$ or without STU cells are capable of learning to better reconstruct the input image. However, without $D$ (w/o $D$) we are not able to properly edit the material appearance since no feedback about such edits exists. On the other hand, concatenating skip-connections directly, without including information from the attribute, hampers the decoder's ability to generate the edited image from the latent code $\mathbf{z}$ (w/o STUs). Figure~\ref{fig:ablation_modules}, depicts how not using a discriminator $D$ generates an image almost equal to the input, not allowing for editing. Besides, not including the STU cell does not convey the desired appearance according to the target high-level attribute in the final image. 
\begin{figure*}[t]
  \centering
  \mbox{} \hfill
    \includegraphics[width=\linewidth]{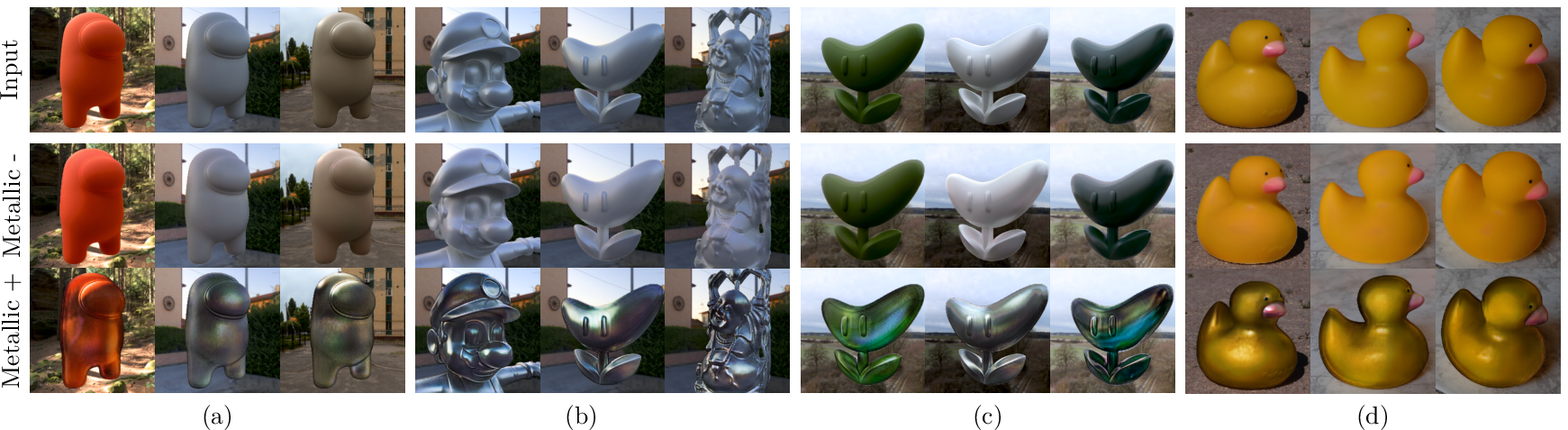}
    \caption{\label{fig:synthetic_bg_metallic} 
    Examples of edited images by our framework. (a) Renders of the Among Us geometry under three different illuminations, and two constant materials. (b) Renders of Mario, Boomerang Flower, and Buddha geometries with the same material and light conditions. (c) Renders of the Boomerang Flower geometry with the same illumination but different materials. (d) Photographs of a rubber duck taken in different places under three light conditions. Our framework is capable of producing compelling and consistent edits in all cases. The ``+'' corresponds to an increase in the target \textit{metallic} attribute value $\mathbf{a t t}_t$, while the ``-'' corresponds to a decrease.}
\end{figure*}

\subsection{Consistency of the Edits}
\label{sec:consistency}

We test the robustness of our framework exploring its editing ability through samples from the evaluation dataset. Since we have a wide collection of rendered images with diverse materials, illuminations, and geometries; we can fix a scene parameter (i.e., illumination, geometry, or material) and vary the other two.
When working with in-the-wild photographs, scene parameters are unknown. We can not modify the material or geometry of the objects but we are able to change their illumination conditions by placing them elsewhere. 
Figure~\ref{fig:synthetic_bg_metallic} (a), (b), (c) shows images edited by our framework for the \textit{metallic} attribute while varying scene parameters for synthetic images. On the other hand, in (d) we vary the illumination conditions for in-the-wild photographs of the same object (for further results see the Supplementary Material). We can see how our framework produces consistent results in the different conditions, and with synthetic and in-the-wild photographs.

We also test the temporal consistency of our framework by editing two video sequences frame by frame for both 
\textit{metallic} and \textit{glossy} attributes. 

\subsection{Comparison with the State of the Art}
\label{sec:comparison}

We compare our results against the method of Delanoy et al.~\cite{delanoy22CGF}. We show results reconstructing the input image using the crowd-sourced perceptual attribute as the input, and editing the input using a different target perceptual attribute. Since the method of Delanoy et al. also needs the normal map as the input, we evaluate in-the-wild photographs with their estimated normal map (using Delanoy et al. estimator), and synthetic images with a perfectly normal map. Our method does not need the normal map as the input. We rely on qualitative and quantitative comparisons employing four metrics: Pixel-to-Signal Noise Ratio (PSNR),  Structural Similarity Index (SSIM), Mean Squared Error (MSE), and Mean Absolute Error (MAE). 

\begin{figure}[t]
    \centering
    \includegraphics[width=\linewidth]{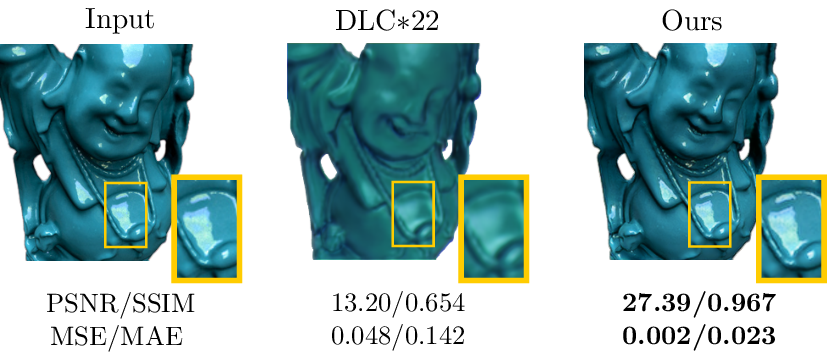}
    \caption{\label{fig:rec_comparison} 
    A demonstration of the reconstruction quality for Buddha geometry. Yellow insets show regions of the object's surface with specular reflections.}
\end{figure}

\paragraph*{Quantitative Evaluation} To evaluate the reconstruction ability we rely on the synthetic images with their crowd-sourced perceptual attributes. Table~\ref{tb:psnr_ssim_pm} shows we outperform the state of the art as a result of introducing STU cells in each skip-connection. As illustrated in Figure~\ref{fig:rec_comparison}, our method keeps high-frequency details from the input image without the need of a normal map of the object's surface as the input. We can see that specular reflections are present on the reconstructed image while the previous method blurs them, keeping only low frequencies. 

\begin{table}[t]
\caption{\label{tb:psnr_ssim_pm} Average PSNR, SSIM, MSE and MAE reconstructing the input image on the synthetic 
dataset. We can see how our framework outperforms the method proposed by Delanoy et al.~\shortcite{delanoy22CGF}.}

\centering
    \begin{tabular}{c c c c c }
    \hline
    \textbf{Method} & PSNR $\uparrow$ & SSIM $\uparrow$ & MSE $\downarrow$ & MAE $\downarrow$ \\ \hline
    \textbf{DLC*22}  &  15.989 & 0.761 & 0.095 & 0.031 \\ 
    \hline
    \textbf{Ours}    &  \textbf{27.388} &	\textbf{0.967}  & \textbf{0.002}
     & \textbf{0.023}      \\ 
    \hline
    \end{tabular}
\end{table}

\begin{figure}[t]
    \centering
    \includegraphics[width=\linewidth]{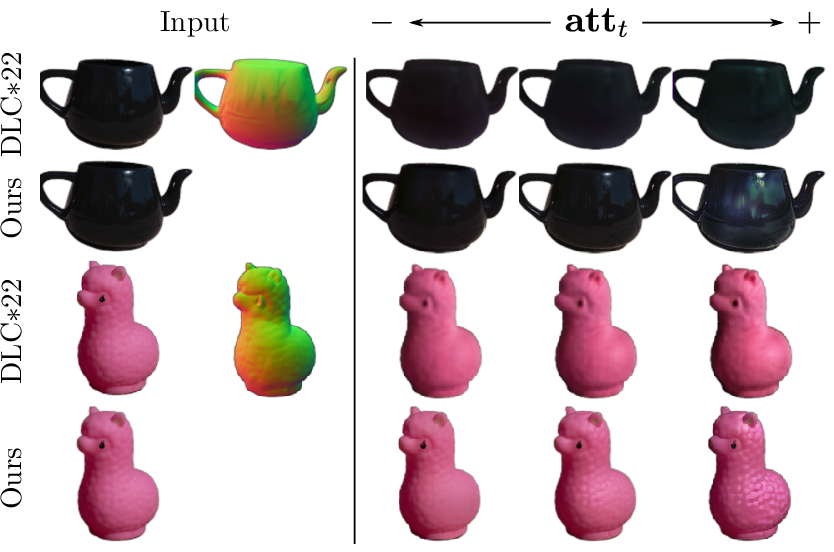}
    \caption{\label{fig:editing_comparison}
    Comparison editing the \textit{glossy} attribute using the method of Delanoy et al.~\shortcite{delanoy22CGF} and our framework for two in-the-wild photographs. Our framework only requires the photograph as the input while the work of Delanoy et al. needs to estimate the normal map. We can see how our method better recovers the \textit{glossy} appearance of the object when edited. Besides, it is able to recover better high-frequency details. The ``+'' and ``-'' indicate whether the target high-level perceptual attribute increased or decreased.}

\end{figure}

\begin{figure}[t]
    \centering
    \includegraphics[width=\linewidth]{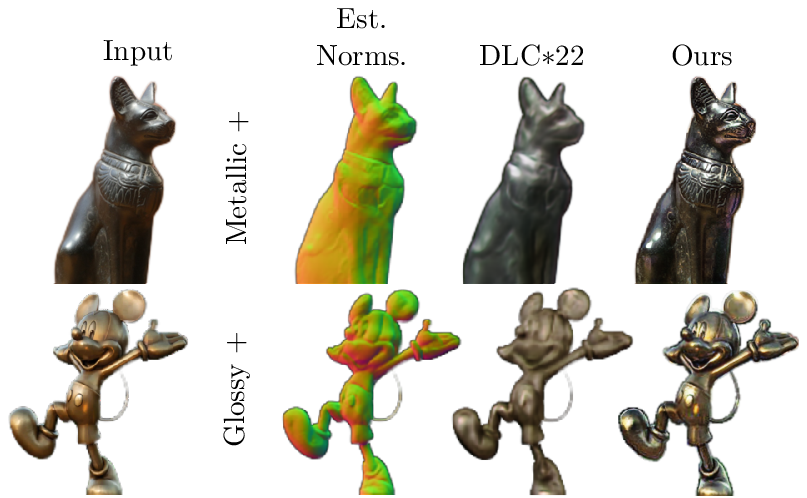}
    \caption{\label{fig:normal_fail} Delanoy et al.~\cite{delanoy22CGF} output is highly dependent to the estimated normal map of the input image.  A low-frequency estimation of the normal map yields dull edits of material appearance. The ``+'' indicates an increase of the target attribute.}
\end{figure}

\paragraph*{Image Editing} In Figure~\ref{fig:editing_comparison} we can see a comparison between the edited images by our method and the one by Delanoy et al.~\cite{delanoy22CGF} (for further comparisons see the Supplementary Material). Our approach learns to edit perceptual cues properly while objects' shape remains unchanged. The material appearance edits from Delanoy et al.~\cite{delanoy22CGF} strongly depend on the shape of their estimated normal map~\cite{lacambra21JJI}. This causes geometry details that are not present in the normal map not to be present in the edited image. Also, an inaccurate estimation of the normal map may deform the original shape, especially in in-the-wild photographs where geometries are usually highly complex (see Figure~\ref{fig:normal_fail}). 

\subsection{Comparison with Physically-based Rendering}

\begin{figure}[t]
    \centering
    \includegraphics[width=\linewidth]{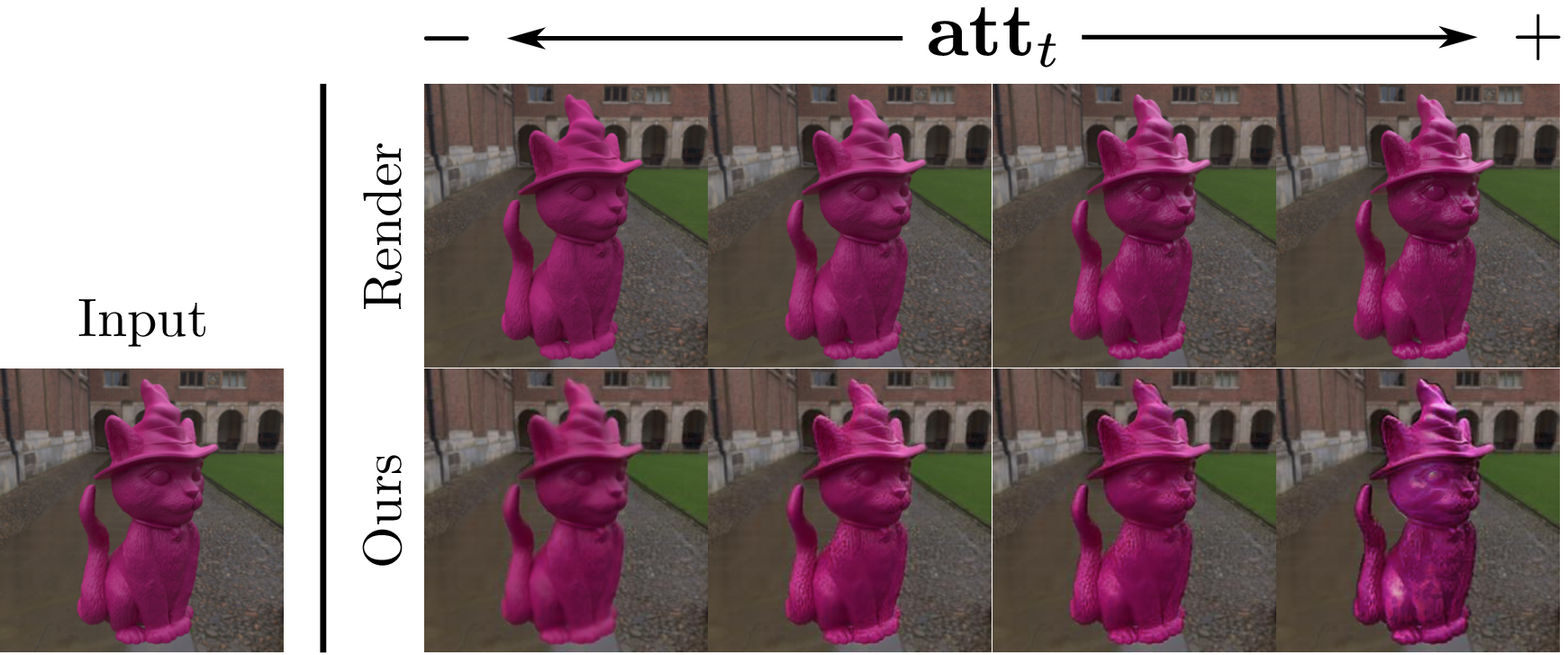}
    \caption{\label{fig:reder_comparison} Results varying the roughness parameter of the Principled BSDF~\cite{burley12SIG,burley15SIG} (top row) and using our framework (bottom row) to edit the \textit{gloss} high-level perceptual attribute. The ``+'' and ``-'' indicate whether the roughness parameter and the target high-level perceptual attribute  increased or decreased.}
\end{figure}

We compare our results with physically-based rendering by 
varying the roughness parameter of the Principled BSDF~\cite{burley12SIG,burley15SIG} in the interval $[0,1]$ and rendering different versions of the same scene (using geometry and illumination not present in our training dataset). The other parameters and the albedo constant and we rely on the physically-based renderer Mitsuba 3~\cite{Mitsuba3} to generate the images.
Then, with our method, we increase and decrease the gloss attribute using the rendered scene with a roughness value of 0.5 as the input. As we can see in Figure~\ref{fig:reder_comparison}, our edits convey the overall appearance of the rendered images. However, we are less accurate when the gloss attribute is increased. This may be explained since removing existing information (highlights) is easier than introducing missing information without providing the environment map.  

\section{Limitations and Future Work}

\begin{figure}[t!]
    \centering
    \includegraphics[width=\linewidth]{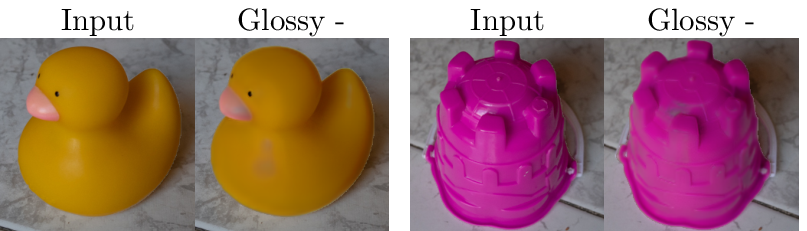}
    \caption{\label{fig:bad_resul} Example of two failure cases of our framework for the \textit{glossy} attribute on photographs. The ``-'' indicates a value of the target attribute set to 0. }
\end{figure}

We have presented and validated a framework for intuitive material appearance editing using single, in-the-wild images. We relied on a large set of images paired with crowd-sourced ratings of high-level perceptual attributes to train our framework. 
We use a generative neural network and devise a loss function that allows us to learn how to edit material appearance based on such high-level attributes, without any pairs of original and edited images. Our results show that the presented method can achieve realistic results, almost on par with real photographs, on a wide variety of different inputs. However, our method is not free of limitations. 
As we can see in Figure~\ref{fig:bad_resul}, when using input photographs that feature highly specular highlights, while able to convey the appearance of the target high-level perceptual attribute, our framework may struggle to edit them. Instead of considering the original albedo to perform edit the highlights when \textit{glossiness} is decreased, the resulting edits feature a dimmed region. 

\begin{figure}[t]
    \centering
    \mbox{} \hfill
    \includegraphics[width=\linewidth]{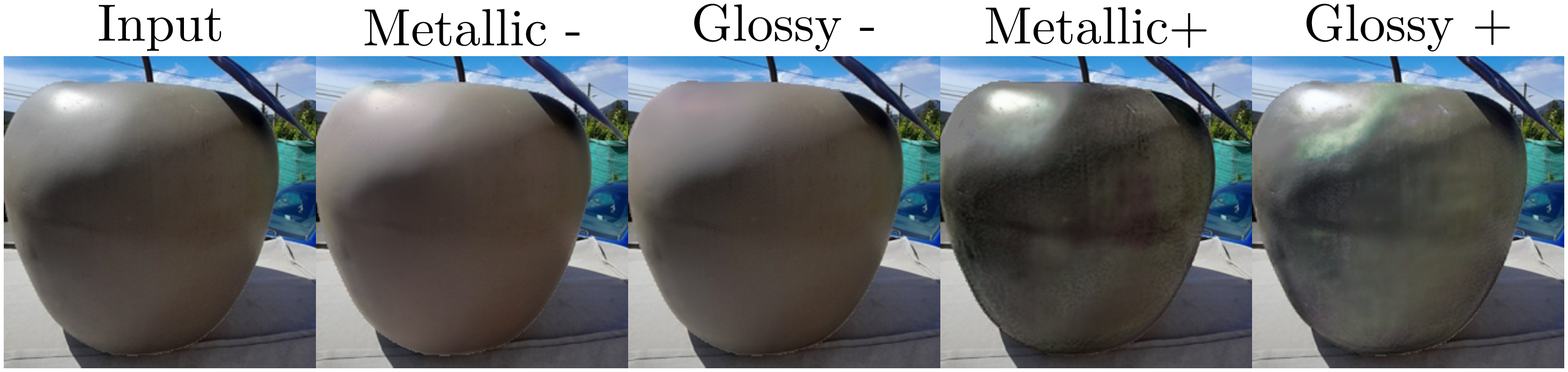}
    \caption{\label{fig:glossy_vs_metallic} Comparison between the \textit{metallic} and \textit{glossy} attribute. The ``+'' indicates that $\mathbf{att}_t$  increases, while ``-'' indicates that it decreases. It is interesting to observe that if the \textit{metallic} attribute is decreased, the gloss is not removed from the image.}
\end{figure}

Material appearance perception and editing pose many challenges that are not fully investigated in this work. 
We propose a data-driven approach for intuitive material editing where we have relied on an existing dataset that was rendered using MERL~\cite{matusik03ACMTG} material measurements. MERL just contains 100 real-world isotropic BRDFs, 
we have tried to increase the variety of such data using different data augmentation strategies during training time. This has allowed us to obtain plausible results, nevertheless, exploring more complex material representations, or including other newly measured material datasets \cite{dupuy18ACMTG,filip14CGF,serrano21SIG} could help obtain more universal models for material editing.
Also, the high-level perceptual ratings in the dataset come from online surveys. While we may identify \textit{metallic} and \textit{glossy} as different properties, there is a certain degree of correlation in user answers between those attributes. In Figure~\ref{fig:glossy_vs_metallic} we show a real photograph edited by our framework. There, we observe that while yielding different results, the increased \textit{glossy} and \textit{metallic} edits share a certain resemblance. Investigating other attributes, or different perceptual data collection strategies may yield improved performance and more intuitive tools. We also observe that the collected perceptual ratings are not distributed uniformly. This result may come from both, a bias in the dataset, and humans’ perceptual behavior. As a result, our framework may not learn to uniformly edit material appearance when varying the target attribute (see Supplementary Material).
Another potential approach to generate the dataset would be to rely on a BRDF model to generate images, and input a parameter of this material model (e.g., roughness) as the target attribute. However, while the framework may be trained using this data, in this work we are addressing the more complex problem of editing appearance from high-level perceptual attributes where the users have factored in all potential confounding factors of perception (including the material model) in their ratings~\cite{lagunas21JoV}.
Last, the generative neural network used in this work has been trained on the limit of our hardware. To use higher-resolution images, one could use the proposed method (and its weights) as a backbone, add additional layers, and fine-tune the model while increasing the resolution size of the input. This could be repeated in an iterative process until we get the desired resolution size~\cite{karras17arxiv}. However, although this is a possible approach, its effectiveness requires further investigation.
We have created a framework that is trained for each attribute. Developing a novel methodology that would allow to manipulate an \textit{appearance vector} can help to have a more comprehensive description of material appearance, and more intuitive edits. In addition to the results we have shown, we hope that our work can inspire additional research and different applications around material appearance.  We will make our code available for further experimentation, to facilitate the exploration of these possibilities.

\section{Acknowledgements}
         
This project has received funding from the Government of Aragon's Departamento de Ciencia, Universidad y Sociedad del Conocimiento through the Reference Research Group "Graphics and Imaging Lab", the European Union’s Horizon 2020 research and innovation program under the Marie Skłodowska-Curie grant agreement No 956585 (PRIME), the CHAMELEON project (European Union’s Horizon 2020, European Research Council, grant agreement No. 682080), and MCIN/AEI 10.13039/501100011033 through Project PID2019-105004GB-I00. We want to thank the Institute for Pure and Applied Mathematics, Universidad Politecnica de Valencia, for the provided computational resources. Also, we would like to thank Daniel Martin and Ana Serrano for proofreading, and Adolfo Munoz, Julio Marco, and Diego Gutierrez for the discussions about the project. Manuel Lagunas' work was done as part of a collaboration with Universidad de Zaragoza and it does not interfere with his position at Amazon.

\bibliographystyle{eg-alpha}  
\bibliography{eg_bib}        


\appendix

\section{STU Details}
\label{apx:stu}

The STU cells are a variant of the GRU~\cite{cho14ACLEM,chung14CoRR} model and allow encoder-decoder Convolutional Neural Networks (CNNs) to retain the relevant information over a long period. Let's say we want to send the feature map of the $l^{t h}$ encoder layer denoted by  $\mathbf{f}^l_{enc}$ to the $l^{t h}$ layer of the decoder. Given a single STU, as is shown in Figure~\ref{fig:architecture_stu}, the hidden state $\mathbf{\hat{s}^{l+1}}$ holds the information from the previous STU cell of the layer $l+1$. This input is used to remove information from $\mathbf{f}^l_{enc}$ an generate the output feature map $\mathbf{f}_t^l$ as is shown in Figure~\ref{fig:architecture_stu}. The hidden state $\mathbf{s^{l}}$ of the cell is calculated and sent to the next layer $l-1$. The hidden state $\hat{\mathbf{s}}^{l+1}$ also has information of the target attribute value:

\begin{equation}
    \hat{\mathbf{s}}^{l+1} = \mathbf{W}_t *_U \left[\mathbf{s}^{l+1}, \mathbf{att}_{t} \right].
\end{equation} 

Where $*_U$ and $[\cdot,\cdot]$ denote the upsampling operation followed by a convolution operation, and the concatenation operator respectively.  The update gate $\mathbf{u}$ helps the cell to determine how much of the past information (from the previous cell) needs to be passed along to the future, while reset gate $\mathbf{r}$ is used to decide how much of the past information to forget. These tensors are computed as:

\begin{eqnarray}
    \mathbf{u}^l =  \sigma \left(\mathbf{W}_u * \left[\mathbf{f}_{enc} , \hat{\mathbf{s}}^{l+1} \right]\right), \label{eq:update_gate}\\
    \mathbf{r}^l = \sigma \left(\mathbf{W}_r * \left[\mathbf{f}_{enc} , \hat{\mathbf{s}}^{l+1} \right]\right).
    \label{eq:reset_gate}
\end{eqnarray}

The sigmoid activation function $\sigma(\cdot)$ is applied to normalize the result between 0 and 1. $\mathbf{W}_u$ and $\mathbf{W}_r$ are the weights matrices updated during training. The hidden state $\mathbf{s}^l$ uses $\mathbf{r}^l$ to store the relevant information from the past and is calculated as follows:

\begin{equation}
    \mathbf{s}^l = \mathbf{r}^l \circ \hat{\mathbf{s}}^{l+1}.
    \label{eq:hidden_state}
\end{equation}

Where $\circ$ expresses the Hadamard product. This operation between $\mathbf{r}^l$ and $\hat{\mathbf{s}}^{l+1}$, determines what to remove from the previous cell. To compute $\mathbf{f}_t^l$; first, nonlinearity is introduced in the form of \textit{tanh} to ensure that the values in the candidate feature map $\hat{\mathbf{f}}_t^l$  remain in the interval $[-1, 1]$:

\begin{equation}
         \hat{\mathbf{f}}_t^l = tanh \left(\mathbf{W}_h * \left[\mathbf{f}_{enc}, \mathbf{s}^l \right]\right).
\end{equation}

Finally, the update gate $\mathbf{u}$ is needed to determine what to collect from the candidate feature map $\hat{\mathbf{f}}_t^l$ and $\hat{\mathbf{s}}^{l+1}$, so we compute the Hadamard product as with Equations~\ref{eq:update_gate} and~\ref{eq:reset_gate}, last, the result is convolved by the weights $\mathbf{W}_h$:

\begin{equation}
     \mathbf{f}_t^l =  \left(1 - \mathbf{u}^l \right) \circ \hat{\mathbf{s}}^{l+1} +\mathbf{u}^l \circ \hat{\mathbf{f}}_t^l.
\end{equation}
\section{Additional Details of Our Architecture}
\label{apx:architecture_details}

\begin{figure}[t]
    \centering
    \includegraphics[width=\linewidth]{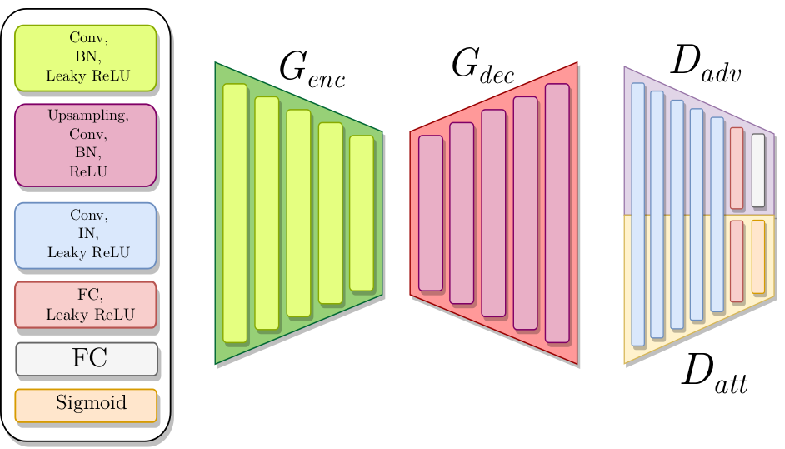}
    \caption{\label{fig:architecture_scheme} Architecture of our generator $G$ and discriminator $D$. BN, IN and FC denote Batch Normalization, Instance Normalization
    and Fully Connected layer respectively. }
\end{figure}

Our framework is composed of an encoder-decoder network $G$ and the auxiliary attribute predictor and image discriminator $D$ only used during training. The generator $G$ is composed of an encoder $G_{e n c}$ made of 5 convolutional layers that reduce the spatial dimensions of the input image by a factor of two, and a decoder $G_{d e c}$ composed of 5 convolutional layers, scaling the input feature of each layer by an upsampling operation. The architecture of the discriminator $D$ is similar to $G_{e n c}$ (5 convolutional layers), but both $D_{a d v}$ and $D_{a t t}$ apply full connected layers to output their predictions, as is shown in Figure~\ref{fig:architecture_scheme}.

\begin{figure}[t]
    \centering
    \mbox{} \hfill
    \includegraphics[width=\linewidth]{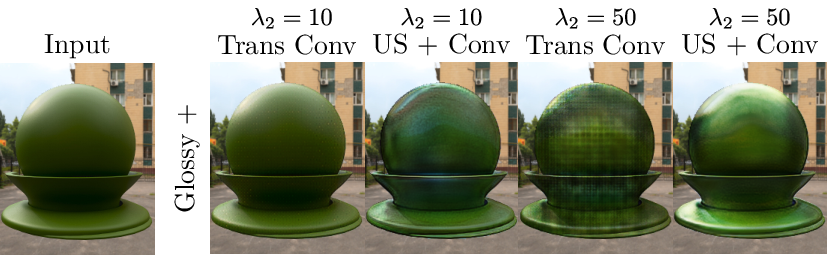}
    \caption{\label{fig:glossy_old_new}
    Ablation studies where we train four versions of our framework.  Arrow (pointing up) indicates we set the target attribute glossy $\mathbf{att}_t$ to 1. From left to right: edited synthetic image by our framework using transposed convolutions and training with $\lambda_2=10$, upsampling followed by a convolutional layer training with $\lambda_2=10$, transposed convolutions, and training with $\lambda_2=50$ and upsampling followed by a convolutional layer training with $\lambda_2=50$.}
\end{figure}

\paragraph*{Trade-off Parameters} Generative models are highly unstable during training, selecting the correct trade-off parameters is crucial. We have observed that the decoder's attribute manipulation loss $\mathcal L_{D_{a t t}}$ decreases rapidly compared to its adversarial loss function $\mathcal L_{D_{a d v}}$. As we see in Figure~\ref{fig:glossy_old_new} increasing $\lambda_2$ improves the editing ability of our framework. On the other hand, the decoder applies several transpose convolutions operations to reconstruct the target image. Unfortunately, this architecture may introduce some artifacts in the edited image. We avoid using transpose convolutions in our framework since they produce artifacts on the final image as is shown in Figure~\ref{fig:glossy_old_new}.

\paragraph*{Comparison with the State of the Art} The number of trainable parameters is an important factor when a deep learning model is being designed because a large number of parameters involves using a lot of memory to train the models and, often, resources are scarce. The work from Delanoy et al.~\cite{delanoy22CGF} propose a framework composed of two generators $G_i$ with $i \in \{1,2\}$ based on a Fader Network~\cite{lample17ANIPS} architecture. $G_1$ compress the low-resolution input image of the single object in a latent code $\mathbf{z}_1$. $G_2$ replicates this behavior but takes high-resolution images as input to generate its latent code $\mathbf{z}_2$. Since both latent codes, $\mathbf{z}_i$ must not contain perceptual information of the high-level perceptual attribute $a$, during training both generators $G_i$ play an adversarial game with a latent discriminator $LD_i$. Also, the authors introduce an image discriminator $D$ that plays an adversarial game with $G_2$ to enhance the quality of edited images. Table~\ref{tb:trainable_parameters_delanoy} shows the trainable parameters of the whole framework.

Our framework is composed of one generator $G$ and one discriminator $D$, while the previous approach needs two generators $\mathcal{G}_i$ helped by other two latent discriminators $\mathcal{L D}_i$ and one image discriminator $\mathcal{C/D}$ to improve the editing ability. In Table~\ref{tb:trainable_parameters_ours} we give the number of total trainable parameters of our framework. Since our framework is simpler than the previous one, it has fewer trainable parameters, so we reduce notably memory usage by reducing the trainable parameters from 58 920 489 parameters to 33 326 314 and thus saving 43\% of memory.

\begin{table}[h!]
\centering
\caption{ \label{tb:trainable_parameters_delanoy} Number of trainable parameters per module for the previous method.}
\begin{tabular}{ c c }
\hline
\textbf{Module}  & \textbf{Trainable Parameters} \\ \hline
$G_1$            & 20 356 515                    \\ \hline
$LD_1$           & 4 326 401                     \\ \hline
$G_2$            & 4 610 179                     \\ \hline
$LD_2$           & 14 813 697                    \\ \hline
$C/D$            & 14 813 697                    \\ \hline
\textbf{Total Parameters} & 58 920 489                    \\ \hline
\end{tabular}

\end{table}

\begin{table}[h!]
\centering
\caption{\label{tb:trainable_parameters_ours} Number of trainable parameters per module for our framework.}
\begin{tabular}{c c}
\hline
\textbf{Module}                         & \textbf{Trainable Parameters} \\ \hline
$G$ ($G_{enc} + G_{dec} + G_{st}$)     & 13 758 280                    \\ \hline
$D$ (both $D_{a t t }$ and $D_{a d v }$) & 19 568 034                    \\ \hline
\textbf{Total Parameters}               & 33 326 314           \\ \hline
\end{tabular}
\end{table}

\end{document}